\documentclass[twocolumn,prb,showpacs,amsmath,amssymb]{revtex4}
\usepackage{graphics,graphpap,amssymb}
\usepackage{graphicx}
\usepackage{hyperref}
\usepackage[dvips]{color}
\begin{document}
\title{Interplay between Kondo tunneling and  Rashba precession}
\author{K. Kikoin$^1$ and Y. Avishai$^2$}
\affiliation{$^1$Raymond and Beverly Sackler Faculty of Exact
Sciences, School of Physics and Astronomy, Tel Aviv University, 69978 Tel Aviv, Israel \\
$^2$Department of physics, Ben Gurion University of the Negev, Beer
Sheva 84105, Israel}
\begin{abstract}
The influence of Thomas -- Rashba precession on the physics of Kondo
tunneling through quantum dots is analyzed. It is shown that this
precession is relevant only at finite magnetic fields. Thomas -- Rashba precession
results in peculiar anisotropy of the effective g-factor
and initiates dephasing of the Kondo tunneling amplitude at low temperature,
that is strongly dependent on the magnetic field.
\end{abstract}
\maketitle

\section{Introduction}

Spin precession due to Rashba coupling \cite{Raby} in a 2D electron
gas (2DEG)  is a specific manifestation of the fundamental Thomas
effect of spin precession in magnetic component of electromagnetic
field due to spin-orbit interaction. This relativistic effect is
strongly enhanced in semiconductors, and in particular in 2DEG in
semiconductor heterostructures.\cite{Rashba} A necessary
precondition for the occurrence of Thomas -- Rashba (TR) precession
in semiconductors is an asymmetry of confinement potential
characterized by a vector $\vec n$ pointing along the electric
field. An interesting physical situation may show up when the TR
precession is noticeable  in 2DEG in which magnetic impurities are
immersed. Since electron scattering by magnetic impurities results
in the Kondo effect, a natural question is whether and how the Kondo
scattering is sensitive to the TR spin precession.  Prima facie it
seems that this precession is irrelevant to the physics of Kondo
screening. In the presence of spin orbit coupling,
 the degenerate two level system is composed of spiral states (Kramers pair), 
 determined  by the spirality winding number (and not by the spin
 projection quantum number as in systems respecting spin rotation invariance). 
 But this distinction simply leads to re-scaling  of the
 Kondo model's parameters without affecting the Kondo physics.
This direct reasoning is supported by basic
arguments\cite{Mewin} stating that, due to time-reversal symmetry,
spin-orbit scattering does not
suppress the Kondo effect  even
though it breaks spin-rotation invariance. Subsequent
investigations\cite{Kavo,Malec} confirmed this conclusion.
At finite external magnetic field, time reversal
invariance is broken, and additional mechanisms affecting Kondo tunneling
arises together with the conventional Zeeman splitting of the impurity
levels, as was demonstrated in Ref. \onlinecite{Mewin} for the case of dirty metals.

On the other hand, it has
been argued\cite{Zar} that an admixture of nonzero angular
modes of spiral states in a 2DEG with  TR precession
\cite{Malec} might cause an enhancement of the Kondo
temperature due to renormalization of the effective exchange
integral.  Similar arguments apply for non-centrosymmetruc cubic
crystals\cite{Agt}. A special case of Kondo effect in the presence of
{\it local} Rashba coupling in quantum wires has recently been considered,
where it is shown\cite{SerSan06,SerSan07,Xu} that
Rashba effect may be the source of resonant states in the bands
and thereby induce the Kondo effect. Thus, there are cases where  
Rashba-type spin-orbit coupling affects the Kondo physics. In that sense we may refer to it as  {\it Kondo-Rashba effect}. \\

In the present paper we discuss the
physical content of the interplay between the TR precession
and the Kondo effect inherent in quantum dots under the constraint of
strong Coulomb blockade.\cite{Kougla} The source of this interlacing 
may be due both to the sizable Rashba-type spin-orbit coupling in the
leads and the TR precession in the complex ring-like
geometry of the dots.\cite{Bergsten,Koenig,Aono}
We stress the specific
features of Kondo-Rashba effect in quantum dot devices in comparison with 
that resulting from magnetic
impurities immersed in
2DEG.\cite{Mewin,Malec} As already noted above, the
TR precession is relevant for Kondo tunneling only under an external
magnetic field. We show here that this relevance stems from the
fact that the spin coordinate axes tilt due to the TR precession.
The tilting axes for the dot and the leads are distinct,  and it is
not possible to match two reference frames in the presence of an external magnetic field.
As a result of the TR effect, the Kondo scattering becomes fully
anisotropic, and this anisotropy is relevant
for the screening mechanism.
In addition, the spatial separation of the Kondo
impurity (the localized electron at the dot) and the leads result in {\it non-local} indirect exchange,
and this non-locality is explicitly related to the TR contribution to the
indirect exchange.

Unremovable mismatch of local magnetic axes is a salient feature
of Dzyaloshinskii-Moriya exchange in some low-symmetry magnetic
crystals.\cite{Mory} It will be shown that the
indirect exchange between spins in the dot and in the leads
mediated by Rashba coupling has the same vector structure as
the Dzyaloshinskii-Moriya interaction between adjacent
localized spins. The relevance TR effect
to indirect exchange  has been perceived in
previous studies. In particular, the
Ruderman-Kittel-Kasuya-Yoshida (RKKY) interaction between
localized spins in 2DEG with Rashba type spin-orbit coupling is characterized by
the above mentioned mismatch of local magnetic
axes.\cite{Kavo,Imam,Simon}. Similar mismatch occurs in
devices consisting of QD with Rashba interaction in  contact with
two ferromagnetic leads\cite{Sun} and in system consisting of two magnetic impurities in a
ring pierced by electric and magnetic fields\cite{Aono}.

\section{Kondo-Rashba coupling
in external magnetic field}

Within the analysis of
Kondo effect in quantum dot with fixed (odd) number of
electrons in weak tunneling contact with source and drain leads, the
starting point is an effective spin Hamiltonian supplemented
by TR term,
\begin{eqnarray}\label{ho}
H &=& \varepsilon_d\sum_\sigma n_{d\sigma} +
\frac{U}{2}\sum_\sigma n_{d\sigma}n_{d\bar \sigma} +
\sum_{k\sigma}\varepsilon_k
n_{k\sigma} \nonumber \\
&+& H_{\rm cot} + H_{{\mbox {\tiny TR}}}.
\end{eqnarray}
The first two terms encode the quantum dot, with electron operators
$d^{}_\sigma, d^\dag_\sigma$, number
operator $n_{d\sigma}=d^\dag_\sigma d^{}_\sigma$, discrete electron level $\varepsilon_d$  and Coulomb blockade energy $U$.  The continuum (band) states in the leads are
characterized by energies $\varepsilon_k$ and
number operators $n_{k\sigma}=c^\dag_{k\sigma} c^{}_{k\sigma}$.
 Assuming the left ($l$) and
right ($r$) leads to be identical, only the even combination
$c_{k\sigma}=(c_{lk\sigma}+c_{rk\sigma})/\sqrt{2}$ survives in the
effective Hamiltonian.
 The next
term, $H_{\rm cot}$, represents an effective cotunneling resulting from the Schrieffer-Wolff (SW) transformation applied on the
original Anderson Hamiltonian. The last term  in (\ref{ho}) stands for the TR precession.

 In order to expose the key features of the interplay between
 TR precession and Kondo tunneling and to elucidate the
triggering role of magnetic field, we first
adopt a phenomenological approach.
Consider a model where both the leads and
the dot are subject to TR precession. Each subsystem $i=l$ (for lead), $d$ (for dot) is
characterized by its own TR coupling with a Rashba vector $\vec
n_i$ and coupling strength $\vec w_i$. (The microscopic substantiation for this
model will be presented at the end of this
section). The effective spin Hamiltonian $H_{\rm s}$ for the
lead-dot device in an external magnetic field $\vec H$ (entering through the the Zeeman
Hamiltonian $H_{\mbox {\tiny Z}}$) has the form
\begin{eqnarray}\label{hg}
&&H_{\rm s} = H_{{\mbox {\tiny TR}}} +  H_{{\mbox {\tiny Z}}} + H_{\rm cot} =   \\
&&\vec n_d \cdot (\vec S \times \vec w_d) + \vec n_l \cdot (\vec
\sigma \times \vec w_l)+ {\vec h}_d \cdot {\vec S}  +{\vec h}_l
\cdot {\vec \sigma}  + J\vec S \cdot \vec \sigma. \nonumber
\end{eqnarray}
Here  $\vec S$ is the dot electron spin 1/2 operator, $\vec \sigma = \sum_{kk'}\sum_{\sigma\sigma'}
c^{\dag}_{k\sigma}\vec \tau c^{}_{k'\sigma'}$ is the lead spin 1/2
conduction electrons operator, $\vec \tau$ is the vector of Pauli
matrices and ${\vec h}_i = g_i\mu_{{\mbox {\tiny B}}}{\vec H}$. The
TR coupling is given by $\vec w_i= \alpha_i \vec p_i$, where
$\alpha_i$ and $\vec p_i$ are TR coupling constants and momentum
operators for dot and lead subsystems.
 External
magnetic field $\vec H$ fixes the direction of the original $z$
axis of spin coordinate system, but the vectors $\vec w_d,~\vec
w_l$ in general case are not parallel and have different moduli.
In many cases the factors $g_i$ are also different in magnitude
and sometimes they even have opposite signs, so we retain the
index $i$ in the Zeeman terms as well.

It is seen from Eq. (\ref{hg}), that the spin precession described
by $H_{\mbox {\tiny TR}}$ results in rotation of spin axes
established by the Zeeman term $H_{\mbox {\tiny Z}}$, but the
rotation angles are {\it different} for dot and lead subsystems,
\begin{equation}\label{spinrot}
\vec S' = {\sf T}(\Theta_d,\Phi_d) \vec S,~~~\vec \sigma' = {\sf
T}(\Theta_l,\Phi_l) \vec \sigma,
\end{equation}
where ${\sf T}(\Theta,\Phi)$ is an appropriate rotation matrix (see below).
In the simplest case where both Rashba vectors are
parallel to the $z$-axis but the coupling constants are different in
magnitude, $\vec n_i=(0,0,1)$, $\vec w_i=(w_{ix}, w_{iy},0)$,
the dot Hamiltonian
\begin{equation}\label{zee}
 H_{\mbox {\tiny Z}} +  H_{\mbox{\tiny TR}}=h_{dz} S_z +
(h_{dx}+ w_{dy})S_x + (h_{dy}-w_{dx})S_y~.
\end{equation}
is transformed to a new spin frame by means of the rotation matrix,
\begin{eqnarray}\label{rots}
&&{\sf T}(\Theta_d,\Phi_d) =\\
&& \left(
\begin{array}{ccc}
 \cos \Theta_d \cos \Phi_d & -\cos \Theta_d\sin \Phi_d & \sin \Theta_d \\
   \sin \Phi_d & \cos \Phi_d & 0 \\
  - \sin \Theta_d \cos\Phi_d & \sin \Theta_d \sin\Phi_d & \cos
  \Theta_d
\end{array}
\right)  \nonumber
\end{eqnarray}
The Euler angles are given by the equations
\begin{equation}\label{Euler1}
\tan\Theta_d = \frac{|w_d|}{h_{dz}},~~~  \tan \Phi_d =
\frac{w_{dy}+h_{dx}}{w_{dx}-h_{dy}}.
\end{equation}
Thus, the quantities $w^2_{d\perp}=w_{dx}^2+w_{dy}^2$
and $h^2_{d\perp}=h_{dx}^2+h_{dy}^2$ define the modulus of a planar
component of an effective magnetic field
$\Delta_\perp^2=w_{d\perp}^2+ h_{d\perp}^2$. Similar
transformation for the Hamiltonian $ H_{\mbox {\tiny Z}}^{(l)}+
H^{(l)}_{\mbox{\tiny TR}}$ yields analogous equations to (\ref{Euler1})
 for the Euler angles
$(\Theta_l,\Phi_l)$,  with $w_l,
h_l$ substituted for $w_d, h_d$.

\begin{figure}[h]\begin{center}
  \includegraphics[width=2.7cm,angle=0]{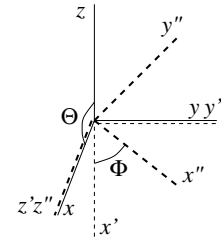}
  \caption{Rotation of spin axes induced at $\vec H=0$ by $H_{\mbox{\tiny TR}}$. The two
  Euler angles are $\Theta=\pi/2$, $\Phi=\pi/4$. Intermediate and final
  coordinates are indicated by primes and double primes, respectively.
  Initial, intermediate and final coordinates are shown by solid,
  dashed and bold dashed lines, respectively.}\label{frot2}
\end{center}
\end{figure}

If the system preserves the square symmetry, then $\Phi_d = \Phi_l
=\pi/4$, but $\Theta_d \neq \Theta_l$ unless $\vec H =0$. In the
latter case $\Theta_d = \Theta_l = \pi/2$, and the rotation of
spin coordinates is the same for both subsystems. Fig. \ref{frot2}
illustrates this rotation.
It follows from (\ref{spinrot}) that, after rotation operation, the cotunneling part of the
spin Hamiltonian (\ref{hg})  acquires the form
\begin{equation}\label{dzmo}
H_{\rm cot}= {\widetilde J} \vec S'(\Omega_d)\cdot
\vec\sigma'(\Omega_l),
\end{equation}
(cf. Refs. \onlinecite{Imam,Sun,Aono}). Here $\Omega_{d
(l)}=\{\Theta_{d (l)},\Phi_{d(l)}\}$. Thus we conclude that the unified spin
coordinate system  for the dot and the leads shown in Fig.
\ref{frot2} may be established only in zero magnetic field $\vec H=0$.
Otherwise, one deals with anisotropic Kondo tunneling, and this
anisotropy is relevant when $\vec H\neq 0$.\cite{Affl}

\begin{figure}[h]\begin{center}
  \includegraphics[width=2.5cm,angle=0]{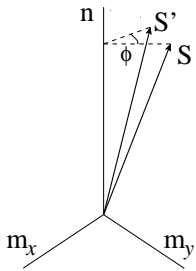}
  \caption{Spin rotation in the presence of the Rashba vector $\vec n$ directed
  along $\hat {\bf z}$. $m_x, m_y$ are the components of the unit vector in the $(x-y)$
  plane.}\label{frot1}
\end{center}
\end{figure}

The indirect exchange Hamiltonian (\ref{dzmo}) may be reduced to
the familiar Dzyaloshinskii-Moriya form in case of strong magnetic
field $h_{iz} \gg w_i$. In this case the angle $\Theta_i \ll
\pi/2$,, so that $\sin \Theta_i \approx w_i/h_{iz},~\cos \Theta_i
\approx 1$. On the other hand, the difference between $\Phi_d$ and
$\Phi_l$ can be substantial, especially when the planar magnetic
field $h^{}_{i\perp}$ is comparable with $w^{}_i$. In this case
the axes $m^{}_{dz}, m^{}_{lz}$ of the spin reference frames for $\vec
S'$ and $\vec \sigma'$ are nearly parallel, whereas the divergence
between the in-plane projections $\{m_{dx},m_{dy}\}$ and
$\{m_{lx},m_{ly}\}$ may be noticeable. One may then choose the frame
$\texttt{M}_l=\{m_{lx},m_{ly},m_{lz}\}$ connected with the leads
as the common reference frame for spins $\vec S$ and $\vec \sigma$
and then expand the rotated spin $\vec S'$ (\ref{spinrot}) around
the spin $\vec S$ determined in the frame \texttt{M}$_l$. Neglecting
the small difference of the projections along $z$, the tilt $\phi$ of the dot spin can
be presented by the vector equality
\begin{equation}\label{1}
\vec S' =  \vec S+\phi \ (\vec n_l \times \vec S),
\end{equation}
(see Fig. \ref{frot1}). Here we used the fact that the Rashba
vector $\vec n_l$ coincides with $m_{lz}$).
The angle $\phi \sim \Phi_{d}-\Phi_l$ is assumed to be small.
Otherwise, a more general expression
\begin{equation} \label{39}
\vec S' =\vec S + \sin\phi ((\vec n \times \vec S))+ \cos \phi
(\vec S - \vec n (\vec n\cdot \vec S))
\end{equation}
should be used instead of (\ref{1}).

Substituting Eq. (\ref{1}) into (\ref{dzmo}), we arrive at the
effective cotunneling Hamiltonian expressed in the reference frame \texttt{M}$_l$ related
to the leads
\begin{equation}\label{dzma}
H_{\rm cot}= J {\vec S}\cdot{\vec \sigma} + {\vec j} (\vec S
\times {\vec \sigma})
\end{equation}
where ${\vec j}= J\phi\vec n$ is the TR induced anisotropic
component of the exchange coupling constant.

A microscopic substantiation of the
phenomenological assumption (\ref{spinrot}) should now be presented. The TR
precession in 2DEG is presented by continuous set of vectors
${\vec w}_l (\vec k)=\alpha\vec k$, where $\vec k$ is the wave
vector in the 2D Brillouin zone.\cite{Malec} To reduce this continuum
to a single vector, one should explicitly take into account the
{\it spatial non-locality} of the lead-dot indirect exchange
induced by cotunneling processes.

In realistic devices, the TR coupling
exists in the planar leads, and there is no generic spin-orbit
interaction in the dot. Then the lead continuum is encoded in
Kondo tunneling through the properties of the band electrons in the point ${\vec R}$,
which denotes the "entrance" coordinate of the tunneling channel
relative to the dot spin position (located at
at $\vec{R}=0$). Taking into account the non-locality of electron cotunneling,
one should write the effective spin Hamiltonian obtained by means
of the SW transformation in the form
\begin{equation}\label{nonloc}
H_{\rm cot} = J \vec S \cdot {\vec \sigma}_{\vec R}
\end{equation}
where $ {\vec \sigma}_{\vec R}=  \sum_{{{\vec k}{\vec
k}'}}\sum_{\sigma\sigma'} c^{\dag}_{{\vec k}\sigma}\vec \tau
c^{}_{{\vec k}'\sigma'}\exp [{i({\vec k}-{\vec k}'){\vec R}}]$.
Then the TR field is presented by its local component in the point
$\vec R$ (see, e.g., Ref. \onlinecite{Imam})
\begin{equation}\label{Rloc}
H_{\mbox{\tiny TR}}^{loc}= \alpha_l ({\vec \sigma}\times \hat r)
F(R).
\end{equation}
 Here $\hat r = \vec R/R$ is a unit
vector along $\vec r$ and $F(R)$ is the form factor arising within the procedure of Fourier
transformation. This means that the planar TR components of
the effective magnetic field in the leads are the components of the
vector ${\vec w_l}= \alpha F(R)\hat {r}$.

Another system where the conjecture expressed in Eq.~(\ref{spinrot})
is realized consists of a quantum dot possessing TR coupling term, whereas the
spin-orbit interaction in the leads is negligible. This regime
may be realized, e.g., in transition metal-organic complex
adsorbed on a metallic substrate in contact with a nano-tip of
a tunneling microscope. In this type of devices the source of the TR term
is the asymmetry of the electric field induced by the nano-tip. Then it
is natural to choose the frame for the leads with the axis
$\sigma_z
\parallel h_z$ and two other axes oriented in such a way that the
system of coordinates $\texttt{M}_d\{m_{dx},m_{dy},m_{dz}\}$ is only
slightly tilted relative to the reference frame. Then one may
adjust the two coordinate systems by means of the vector equality
\begin{equation} \label{39b}
\vec S' =\vec S + \phi ((\vec n_d \times \vec S))
\end{equation}
like in Eq. (\ref{1}) and thereby arrive at the same anisotropic spin
Hamiltonian (\ref{dzma}), which describes the interplay between TR
and Kondo mechanisms.

\section{Scaling analysis of anisotropic Thomas-Rashba-Kondo Hamiltonian}

Based on the above analysis, we now study the
interplay between the TR precession and the Kondo effect in the weak
coupling regime $T\gg T_K$, where the RG scaling approach for
identifying the fixed points is applicable. In the two limiting cases of
strong and weak magnetic field, the  general Hamiltonian (\ref{dzmo})
is reducible to a simplified effective Hamiltonian (\ref{dzma}), as discussed below.

\subsection{Strong magnetic field}

Following our analysis of the previous section,
the anisotropic Thomas-Rashba-Kondo Hamiltonian reads,
\begin{equation}\label{30}
H=\sum_{k\sigma}\varepsilon_k n_{k\sigma}  + \vec h_d \cdot \vec S
+  J {\vec S}\cdot{\vec \sigma} + {\vec j} (\vec S \times {\vec
\sigma}).
\end{equation}
This form entails a Rashba vector that is parallel to the
$z$-axis and a strong external magnetic field $\vec H\parallel \hat {\bf z}$, so
that $\vec h_d = \{w_{dx}, w_{dy}, h \}$ and $h\gg w_d$.  The
second term is the spin Hamiltonian of the isolated dot
\begin{equation}\label{zee3}
 H_{{\mbox {\tiny Z}}}+  H_{\mbox{\tiny TR}}=h_z S_z
+\frac{1}{2}\left( w_dS^+ + w_d^*S^-\right)~,
\end{equation}
where $w_d= w_{dy}+iw_{dx}$. The last two terms in Eq. (\ref{30})
form the co-tunneling part, rewritten as,
\begin{equation}\label{7b}
H_{{\mbox {\tiny cot}}}=\frac{1}{2}\left(J_-\sigma_+S_- +
J_{+}\sigma_-S_+\right) + J \sigma_z S_z,
\end{equation}
with
\begin{equation}\label{15}
J_{\pm}=J(1\pm i\phi), 
\end{equation}
and $\phi\approx |w_d|/h$. Thus, the spin-related part of the above
Hamiltonian is generically anisotropic. To expose the evolution (flow) of
the anisotropy parameters we define
\begin{equation}\label{16}
J_+ -J_- = 2iJ\phi \equiv 2ij_{\mbox {\tiny TR}}~,
\end{equation}
where $j_{\mbox {\tiny TR}}=J\phi_{d}$ is the modulus of the Kondo -- Rashba vector
coupling in the Hamiltonian (\ref{30}). It is readily seen from Eq.~(\ref{16})
that the magnetic anisotropy induced by TR precession {\it
increases} on approaching the standard infinite fixed point and
hence it is relevant.

In the weak coupling limit one may study the Kondo problem using
 "poor man's scaling" perturbative approach.\cite{Andrg}
In our case with TR term present, deviation from the standard scaling paradigm arises already in
zero order in the exchange constant because the Kondo problem should be
solved in the presence of an effective "magnetic" field given by Eq.
(\ref{zee3}). Using the pseudofermion representation for spin
operator $\vec S = \sum_{\sigma\sigma'} f^\dag_\sigma \vec \tau
f^{}_{\sigma'}$, we rewrite (\ref{zee3}) as
\begin{equation}\label{20}
H_{{\mbox {\tiny TR}}} +  H_{{\mbox {\tiny
Z}}}=\frac{h}{2}(f^\dag_\uparrow f^{}_\uparrow - f^\dag_\downarrow
f^{}_\downarrow ) +\frac{1}{2}\left(w_d f^{\dag}_\uparrow
f^{}_\downarrow + w_d^*f^{\dag}_\downarrow f^{}_\uparrow \right)
\end{equation}
In accordance with the arguments adduced in the previous section,
this "zero-order" Hamiltonian cannot be diagonalized by means of
rotation of spin coordinate frame. Therefore the bare Matsubara
spin-fermion propagators $g_{\sigma\sigma'}(\tau)= -\langle T_\tau
f^{}_\sigma(\tau)f^{\dag}_{\sigma'}(0)\rangle$ and their Fourier
transforms $g_{\sigma\sigma'}(\varepsilon)$ form a $2\times 2$
matrix
\begin{equation}\label{19}
\hat {g}= \left(
\begin{array}{cc}
  g_{\uparrow\uparrow}(\varepsilon) & g_{\uparrow\downarrow}(\varepsilon) \\
  g_{\downarrow\uparrow}(\varepsilon) & g_{\downarrow\downarrow}(\varepsilon) \\
\end{array}%
\right).
\end{equation}
Here
\begin{eqnarray}\label{21}
&&g_{\sigma\sigma}=(\varepsilon-\bar\sigma h/2)/(\varepsilon^2-\Delta^2), \\
&&g_{\uparrow\downarrow}=w_d^*/2(\varepsilon^2-\Delta^2),~~
g_{\downarrow\uparrow}= w_d/2(\varepsilon^2-\Delta^2), \nonumber
\end{eqnarray}
$\varepsilon$ is the Matsubara  frequency, $\Delta=\sqrt{h^2+
|w_d^2|}/2$ is the modulus of the effective magnetic field, including
the contribution of TR precession. Both Zeeman components
contribute to each of these functions. In the spinor representation
the pseudofermion propagator may be represented as a combination of
"normal" (spin conserving) and anomalous term \cite{foot1}
\begin{equation}\label{21b}
\hat g = \hat g^{}_\|+\hat g^{}_\perp\equiv g_0 S_z + g_1 \vec
n_d\cdot (\vec S\times \vec w_d).
\end{equation}
(explicit form of for $g_0$ and $g_1$ is easily derived from
(\ref{21})).

\begin{figure}[h]
\includegraphics[width=5cm,angle=0]{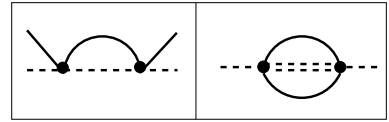}
\caption{Left panel: diagram contributing to scaling equations in
a single-loop leading logarithm approximation. Right panel:
leading logarithmic correction to the spin-fermion self energy in
the limit of strong magnetic field $h\gg |w_d|$. Dashed and
double-dashed lines stand for the longitudinal and transversal
components of spin-fermion propagator (\ref{19}) solid lines
correspond to electron propagators, vertices $J$ are denoted by
circles.} \label{f.2}
\end{figure}

The scaling equations for the Kondo effect derived in a
single-loop approximation Fig.
 \ref{f.2} acquire the following form
\begin{eqnarray}\label{17}
\frac{dJ_\|}{d\eta} = -J_+J_-,~~~ \frac{dJ_\pm}{d\eta} = - J_\pm
J_\|
 \end{eqnarray}
Here and below we turn to dimensionless coupling constants $J \to
\nu_0J$ etc, where $\nu_0 \sim D_0^{-1}$ is the electron density
of states in the leads assumed to be constant in the vicinity of
the Fermi level. The scaling variable is defined as $\eta=\ln
(D/D_0)$.

 Unlike the standard flow equations,\cite{Andrg} the
transverse  components of the exchange parameters are complex
(\ref{15}). With the help of (\ref{16}) we transform (\ref{17}) into
\begin{eqnarray}\label{flow}
\frac{dJ}{d\eta} &=& -(J^2 + j_{\mbox {\tiny TR}}^2)  , ~~~
\frac{dj_{\mbox {\tiny TR}}}{d\eta} = - J j_{\mbox {\tiny TR}}
\end{eqnarray}
The second equation describes the evolution of the imaginary TR
correction to the transverse part of the exchange vertex. Here and
below the index $0$ labels the initial scale of the energy and
coupling parameters of the Hamiltonian (\ref{30}).

Integration of Eqs. (\ref{flow}) with the boundary conditions
$J(0)=J_0,~j_{\mbox {\tiny TR}}(0)=j_0$ gives (within
logarithmic accuracy)
\begin{equation}\label{23}
J(\eta)= \frac{\widetilde J_0}{1-\widetilde J_0\eta},~~~
 j_{\mbox {\tiny TR}}=
 \frac{j_0}{1-\widetilde J_0\eta}.
\end{equation}
Here $\widetilde J_0 = \sqrt{J_0^2 + j_{\mbox {\tiny TR}}^2}$.
This result means that although the imaginary TR component of
the exchange anisotropy increases with reduction of the energy scale,
its contribution to the real longitudinal parameter $J(\eta)$
results only in the enhancement of the Kondo temperature from $T_K =
\exp(-1/J_0)$ to $\widetilde T_K = \exp(-1/\widetilde J_0)$ and
does not influence the fixed point. One should, of course,
remember that the Kondo resonance is in fact split by the effective
magnetic field $\Delta$ entering the poles of the spin-fermion
propagators (\ref{21}), where the axial component of this field
$\sim |w_d|$ arises due to the TR precession.

The effective field $\Delta$ is also affected by the interplay
between Kondo tunneling and TR precession. Whereas only the
diagonal part $\hat g^{}_\|$ of the bare spin propagator
(\ref{21b}) contributes to the system of scaling equations (\ref{flow}), the
transverse component $\hat g^{}_\perp$ renormalized by Kondo
co-tunneling enhances the planar component of the effective magnetic field.
In the limit $|w_d|\ll h$ the off-diagonal spin-fermion propagator
has the form
\begin{eqnarray}\label{27}
&&\hat{g}^{}_{\perp}(\varepsilon)\approx \frac{S^+w_d +
S^-w_d^*}{2h}\cdot g_1(\varepsilon)\\
&& g_1(\varepsilon)= \left(\frac{1}{\varepsilon-\Delta}-
\frac{1}{\varepsilon+\Delta}\right)\nonumber
\end{eqnarray}
Since there is no counterpart to this propagator in the Green
functions of band electrons, the Kondo loops in the self energies containing $\hat
g^{}_\perp$ generate extra terms, which do not conserve spin, namely contain the factors
$S^\pm\sigma_z+ S_z\sigma^\pm$. The corresponding diagrams are
shown in the right panel of Fig. \ref{f.2}.

\begin{figure}[h]
\includegraphics[width=5cm,angle=0]{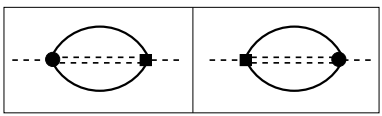}
\caption{TR corrections to the self energies of spin-fermion
propagators in the limit of weak magnetic field $h\ll |w_d|$.
Circles and squares denote the vertices  $\sim J$, which
conserve and do not conserve the total spin, respectively (see the
text for details.)}
\label{f.3}
\end{figure}

 The "anomalous" transverse propagators 
$\hat{g}^{}_{\perp}$ defined in Eq.~(\ref{27})
 are responsible for rescaling  the axial
components of the magnetic field $\vec h$ (\ref{zee3}). The
 lowest order diagrams contributing to the self energy of $\hat
g^{}_\perp$ are shown in Fig. \ref{f.3}.
\ \\
The explicit expressions for the self energy diagrams 
for $\hat{g}^{}_{\perp}(\varepsilon)$ and $\hat{g}^{}_1(\varepsilon)$ are, 
\begin{eqnarray}\label{24}
&&\Sigma_{\perp}(\varepsilon)=\frac{S^+w_d +
S^-w_d^*}{8h}\cdot \Sigma_1(\varepsilon)\\
&&\Sigma_1(\varepsilon)=J^2T^2\sum_{\omega_1\omega_2,k_1k_2}G_{k_1}(-\omega_1)G_{
k_2}(\omega_2)g_1(\varepsilon+\omega_1+\omega_2)
    \nonumber
\end{eqnarray}
Here $G_k(\omega)$ are the bare propagators for conduction electrons
in the leads. In terms of real frequencies we find that the
leading logarithmic term of this self energy has the form
\begin{equation}\label{25}
{\rm Re}\Sigma_1(\varepsilon)=2(\varrho_0J)^2h
\ln\frac{D_0}{\max\{(|2\Delta-\varepsilon|), T\}},
\end{equation}
(cf. similar estimates for self energies of spin fermion
propagators describing singlet-triplet configuration in quantum
dots with even occupation at finite bias \cite{KKM}). The
imaginary part of the self energy ${\rm
Im}\Sigma_1(\varepsilon)\sim J^2T$ is irrelevant. Inserting these
estimates in (\ref{24}), we find that this self energy enhances
both real and imaginary parts of the planar magnetic field. At
$T>|2\Delta-\varepsilon|$ corrections to the planar components of
the effective magnetic fields may be estimated as
\begin{eqnarray}\label{31}
\frac{\delta w_{d}}{|w_d|}\sim (\rho_0J)^2\ln\frac{D_0}{T}.
\end{eqnarray}

Thus, we have found that the magnetic anisotropy induced by the TR
precession is enhanced due to the interplay between this precession
and the Kondo co-tunneling. In the limit of strong field $h\gg w_d$
this enhancement acquires the form of "dynamical" contribution to
the planar magnetic field. This "random" field reminds the effect
of exchange anisotropy induced by an edge spin coupled to an open
spin-one-half antiferromagnetic Heisenberg chain.\cite{FraZ} The
Kondo-induced component of the planar field is weak at $T \gtrsim
T_K$, $ |\delta w|_\perp/T_K \sim j\Delta$. However, it generates
its own energy scale
\begin{equation}
T^* = T_K \exp\left(-\frac{1}{ j}\right)
\end{equation}
where the precession induced magnetic field becomes comparable
with the static magnetic field $\Delta$.

\subsection{Weak magnetic field}

In the limit of weak field (or strong TR interaction), namely,  $w_d\gg h$,
the general phenomenological analysis of Section II points toward another
way to  arrive at a Dzyaloshinskii-Moriya form for the TR corrections to the
effective exchange Hamiltonian. Let us consider a model with
nonlocal exchange between the dot and the leads with the effective
exchange given by the Hamiltonian (\ref{nonloc}) in the absence of
TR precession. Assume that the Rashba vectors are parallel in both
lead and dot systems, $\vec n_l \parallel \vec n_d \parallel z$ but $w_d\neq
w_l$. Then the spin Hamiltonian acquires the form (\ref{dzmo}).

In accordance with the kinematic scheme of Fig. \ref{frot2}, at zero
magnetic field and square $(xy)$ symmetry, the Euler angles are
$\Theta_l=\Theta_d=\pi/2$ and $\Phi_l=\Phi_d=\pi/4$. The difference
between the coordinates $(x'',y'',z'')$ for lead and dot spins at
small $h$=$h_z$ is proportional to the deviation of $\Theta_d$ and
$\Theta_l$ from $\pi/2$, namely $\pi/2 -\Theta_i$=$\varphi_i$, where
$\varphi_i\approx h/w_i$. Then the mismatch between the directions
of the vectors $\vec S$ and $\vec \sigma_R$ is small like in Fig.
\ref{frot1}, but the axis $\vec n$ is directed along the coordinate
$z''=x$ of Fig. \ref{frot2}. Returning to the original frame
$(x,y,x)$ we write the bare dot spin Hamiltonian in the form
(\ref{zee3}), and matching  the angles $\Omega_d \to \Omega_l$ means
applying the transformation
\begin{equation}\label{ll}
\vec S' =  \vec S+\varphi \ (\vec n \times \vec S),
\end{equation}
where $\varphi=|\varphi_d-\varphi_l|$ and only the $x$-component
of the vector product survives. The TR correction to the exchange
Hamiltonian acquires the form
\begin{equation}\label{veranom}
\delta H_{\rm cot}= \frac{i\varphi
J}{2}\left[S_z(\sigma^--\sigma^+) + (S^--S^+)\sigma_z \right]
\end{equation}
In this limit the main contribution to the spin-fermion
propagators (\ref{19}) is given by the off-diagonal components
$g_{\sigma\bar\sigma}$, while the residues of the longitudinal
components $g_{\sigma\sigma}$ contain small parameter $\varphi$.
Thence the``anomalous" contribution to the Kondo loops (Fig.
\ref{f.3}) gives the leading contribution to the scaling equations
for the vertices $i\kappa=i\varphi J$ (\ref{veranom}),
\begin{equation}
\frac{\partial \kappa}{\partial \eta} = - \kappa J~.
\end{equation}
which implies scaling evolution of $\kappa$ similar to that of
$j_{\mbox {\tiny TR}}$ (\ref{flow}), (\ref{23}). Then we get an expression for
 the longitudinal component of the self energy
given by the diagrams depicted in Fig. \ref{f.3}:
\begin{eqnarray}\label{28}
\Sigma_\|(\varepsilon)=\frac{i\varphi(w_d-w_d^*)S_z}{4}\Sigma_1(\varepsilon).
\end{eqnarray}
As in Eq. (\ref{25}), the logarithmic renormalization arises in
the self energy for real frequencies, and the magnetic field
enhancement can be estimated similarly to (\ref{31})
\begin{equation}\label{hfield}
\frac{\delta h}{h}\approx
(\rho_0J)^2\ln\frac{D_0}{\max\{|2\Delta-\varepsilon|,T\}}.
\end{equation}

Thus we have found that the interplay between the Kondo scattering
and the TR precession in case where the Rashba vector is parallel
to a magnetic field results in logarithmic enhancement of the planar and
the $z$-component of the effective magnetic field in the limits of strong
and weak external field, respectively. This interplay disappears in
zero field in agreement with the general symmetry
considerations.\cite{Mewin}

\section{Conclusions}

The main result of our analysis of the kinematics of Rashba effect
in a system 'quantum dot plus metallic reservoir' is the statement
that the TR precession in one subsystem is "exported" to another
subsystem by the tunneling processes. This means that the TR
precession always exists both in the dot and in the leads, and the
inequality $\Theta_d \neq \Theta_l$ for the Euler angles related
to the quantization axes in the two subsystems [see Eqs. (\ref{spinrot}) -
(\ref{dzmo})] arises in an external magnetic field, so that the
magnetic quantization axes are never matched. In the limits of
strong and weak magnetic field the Hamiltonian (\ref{dzmo}) is
reducible to the Dzyaloshinsky -- Moriya like form. This conclusion is
quite general, and one may expect similar mismatch in complex
quantum dots, where each constituent dot will be characterized by
its own set of Euler angles $\Omega_{di}$.

As to the physical manifestations of the interplay between
Kondo tunneling and Thomas -- Rashba precession, the main effect
is the sharp anisotropy of the g-factor due to the influence of
the precession on the direction of the effective field $\vec h_d$ [see,
e.g., Eq. (\ref{zee3})]. Due to the contributions of Kondo
processes, this effect is temperature dependent and may be quite
noticeable in case of weak magnetic field (\ref{hfield}).

We restricted our study for the case of local TR effect in the leads
(\ref{Rloc}). The theory may be generalized for the case of
"itinerant" quantization axis following the rotation of the
quantization axis in the 2D Brillouin zone. In this case the
higher angular harmonics of the electron states in the
leads\cite{Malec} should be involved.
\medskip\\

\noindent
{\bf Acknowledgement} The authors are grateful to M.N. Kiselev, A. Nersesyan and A.A. Zvyagin
 for valuable comments. Discussions with Y. Oreg at the initial stage of
this work are highly appreciated.The research of Y.A is partially
supported by an ISF grant 173/2008.


\begin{thebibliography}{99}
\bibitem{foot1} The analog of Eq. (\ref{21b}) is the local Green function $G(\vec R,
\varepsilon)$ for lead electrons in presence of TR splitting,
which consists of two terms, $ G(\vec R, \varepsilon) = G_0(\vec
R, \varepsilon)\sigma_0 + G_1(\vec R, \varepsilon)({\vec
n}_l\times \vec R)\cdot \vec \sigma $. Both conduction subbands
contribute to "normal" and "anomalous" Green function
components.\cite{Imam}
\end{thebibliography}
\end{document}